\documentclass[aps,prl,twocolumn,showpacs,preprintnumbers,amsmath,amssymb]{revtex4-1}
\usepackage{braket}

 \def\ep{{\epsilon}}

 \def\frac#1#2{{#1\over #2}}

 \def\s{\sqrt}

\def\be{\begin{equation}}
\def\ee{\end{equation}}
\def\ba{\begin{eqnarray}}
\def\ea{\end{eqnarray}}

 \def\de{\partial}

 \def\f {\frac}
 \def\ti{\tilde}

 \def\ddd{\cdot\cdot\cdot}
 \def\no{\nonumber \\}

 \def\la{\langle}
 \def\lb{\rangle}
 \def\ep{\epsilon}

\usepackage{color}
\usepackage{graphicx}
\usepackage{dcolumn}
\usepackage{bm}

\begin{document}

\title{Gravity Dual of Quantum Information Metric}
YITP-15-62; IPMU15-0119
\author{Masamichi Miyaji$^{a}$, Tokiro Numasawa$^{a}$, Noburo Shiba$^{a}$,  \\
Tadashi Takayanagi$^{a,b}$ and Kento Watanabe$^{a}$}

\affiliation{$^a$Yukawa Institute for Theoretical Physics,
Kyoto University, \\
Kitashirakawa Oiwakecho, Sakyo-ku, Kyoto 606-8502, Japan}

\affiliation{$^{b}$Kavli Institute for the Physics and Mathematics
 of the Universe,\\
University of Tokyo, Kashiwa, Chiba 277-8582, Japan}

\date{\today}

\begin{abstract}
We study a quantum information metric (or fidelity susceptibility) in conformal field theories with respect to a small perturbation by a primary operator. We argue that its gravity dual is approximately given by a volume of maximal time slice in an AdS spacetime when the perturbation is exactly marginal. We confirm our claim in several examples.
\end{abstract}

\maketitle

{\bf Introduction}

The microscopic understanding of black hole entropy in string theory by 
Strominger and Vafa \cite{SV} implies that quantum information plays a crucial role to understand gravitational aspects of string theory. Indeed, quantum information theoretic considerations have provided various useful viewpoints in studies of AdS/CFT \cite{Ma} or more generally holography \cite{Hol}. Especially, the idea of quantum entanglement has turned out to be crucially
involved in geometries of holographic spacetimes, as typical in the non-trivial
topology of eternal black holes \cite{MaE}. To quantify quantum entanglement
we can study the holographic entanglement entropy \cite{RT}, which is given by the
area of codimension two extremal surfaces.  In the AdS/CFT, this area is equal to the entanglement entropy in conformal field theories (CFTs).

\begin{figure}
  \centering
  \includegraphics[width=5cm]{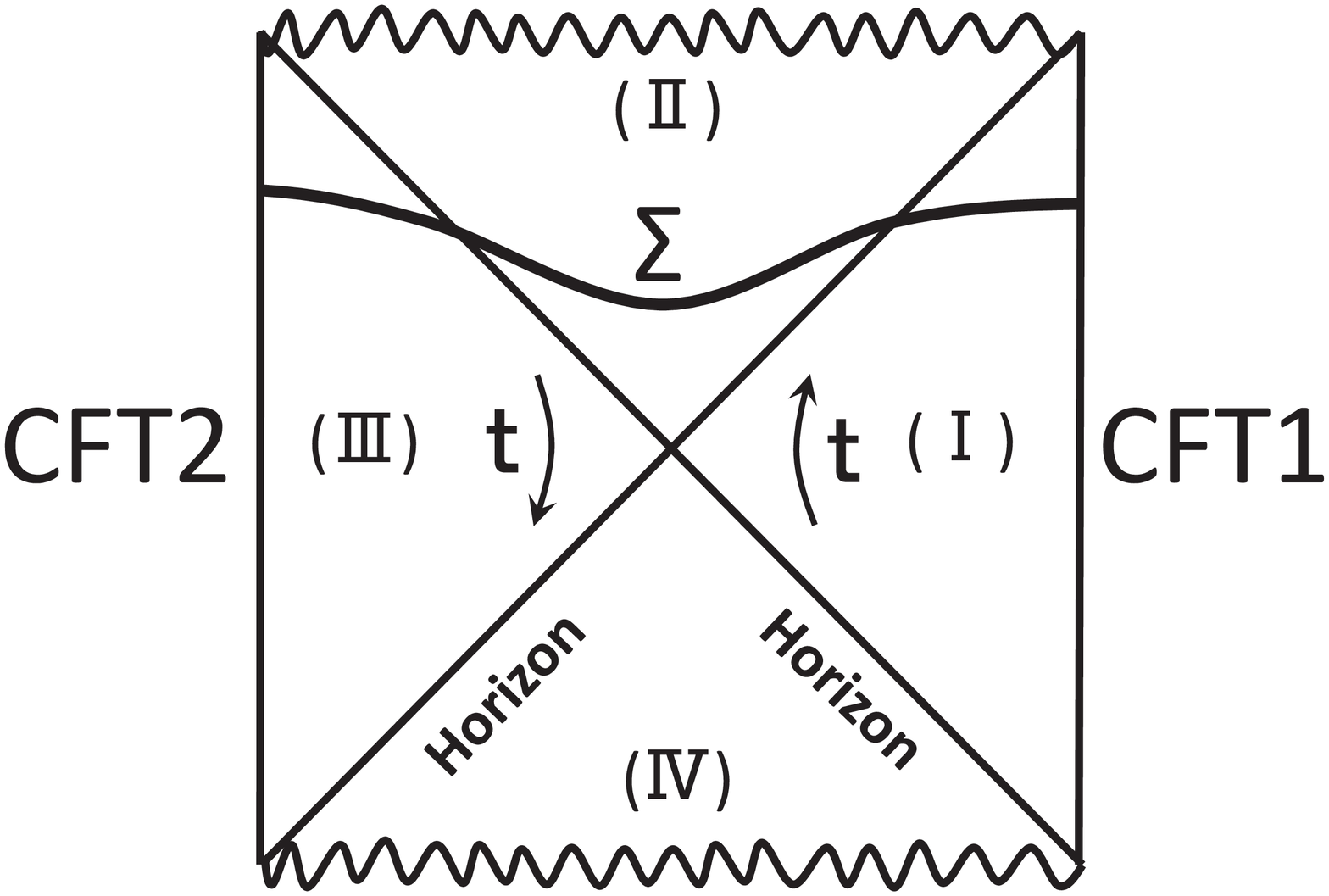}
  \caption{A time slice in the Penrose diagram of eternal AdS black hole which connects the two boundaries dual to the thermofield doubled CFTs.}
\label{fig:pe}
  \end{figure}

It is natural to wonder if there might be some other information theoretic quantities
which are useful to develop studies of holography. As pointed out by Susskind in \cite{Su}
(see also \cite{StSu}), it is also intriguing to find a quantity in CFTs which is dual to a volume of a codimension one time slice in AdS. The time slice can connect two boundaries dual to the thermofield doubled CFTs, through the Einstein-Rosen bridge (see Fig.\ref{fig:pe}). In \cite{Su}, it is conjectured that this quantity is related to a measure of complexity.

The main purpose of this letter is to point out a quantum information theoretic quantity which is
related to the volume of a time slice. This quantity is called quantum information metric
or Bures metric (see e.g.\cite{Bres}), which we will simply call the information metric. Here we mainly consider the information metric for pure states, though it can be defined for mixed states. Consider one parameter family of quantum states $|\Psi(\lambda)\lb$ and perturb $\lambda$ infinitesimally $\lambda\to \lambda+\delta\lambda$. Then $G_{\lambda\lambda}$ is simply defined from the inner product between them as follows:
\be
|\la \Psi(\lambda)|\Psi(\lambda+\delta\lambda)\lb|=1-G_{\lambda\lambda}\cdot (\delta\lambda)^2+O((\delta\lambda)^3).
\label{infmet}
\ee
 This metric  measures the distance between two infinitesimally different quantum states. Since the left-hand side of (\ref{infmet}) is called the fidelity, $G_{\lambda\lambda}$ is also called the fidelity susceptibility (see e.g. the review \cite{Fid} for applications to quantum phase transitions.).

We will argue that $G_{\lambda\lambda}$ when a $d+1$ dimensional CFT is deformed by an
exactly marginal perturbation, parameterized by $\lambda$, is holographically estimated by
\be
G_{\lambda\lambda}=n_{d}\cdot \f{\mbox{Vol}(\Sigma_{max})}{R^{d+1}},  \label{volume}
\ee
 where $n_{d}$ is an $O(1)$ constant and $R$ is the AdS radius. The $d+1$ dimensional space-like surface $\Sigma_{max}$ is the time slice with the maximal volume in the AdS which ends on the time slice at the AdS boundary(ies). See also \cite{MERA} for some other holographic interpretations of information metric.\\

{\bf Information Metric in CFT$_{d+1}$}

Now we introduce the information metric for quantum states in CFTs on $R^{d+1}$, whose
Euclidean time and space coordinates are denoted by $\tau$ and $x$.
We consider the inner product $\braket{\Omega_1|\Omega_2}$ between two states
$\ket{\Omega_1}$ and $\ket{\Omega_2}$. $\ket{\Omega_i}\ (i=1,2)$ are ground states
for the two Hamiltonians $H_i\ (i=1,2)$. We define their Euclidean lagrangians by
$\mathcal{L}_i\ (i=1,2)$ and their partition functions $Z_i\ (i=1,2)$. The inner product is
described by the path-integral:
\ba
&& \braket{ \Omega_2  | \Omega_1 }\no
&& = (Z_1Z_2)^{-1/2} \int D\phi
\exp[-\int d^dx (\int_{-\infty}^{0} d\tau  \mathcal{L}_1 + \int_0^{\infty} d\tau  \mathcal{L}_2  )] .\no
\label{overlap 1}
\ea
Assume the difference $\mathcal{L}_2 -\mathcal{L}_1$
 is infinitesimally small and is written by using the primary operator $O(\tau,x)$
as
\be
\mathcal{L}_2 -\mathcal{L}_1 \equiv \delta \mathcal{L}=\delta \lambda\cdot O(\tau,x).
\label{pert}
\ee

Next, we rewrite (\ref{overlap 1}) by using the expectation value $<\ddd>$
in the vacuum state $\ket{\Omega_1}$:
\begin{equation}
\braket{\tilde{\Omega}_2(\epsilon)| \Omega_1 }
= \dfrac{ \la\exp [ -\int_{\ep}^{\infty} d\tau \int d^dx \delta \mathcal{L}] \lb}
{ (\la\exp [ -(\int_{-\infty}^{-\ep}+\int_{\ep}^{\infty})
d\tau \int d^dx \delta \mathcal{L}] \lb)^{1/2}},
\label{overlap 2}
\end{equation}
where $\delta \mathcal{L} \equiv \mathcal{L}_2 -\mathcal{L}_1$. Here we introduced the UV regularization $\ep$ by replacing the ground state $\ket{\Omega_2}$
with
\begin{equation}
\ket{\tilde{\Omega}_2(\epsilon)} \equiv
\dfrac{e^{-\epsilon H_1} \ket{\Omega_2}}{(\bra{\Omega_2} e^{-2\epsilon H_1} \ket{\Omega_2})^{1/2}}.
\end{equation}
By performing perturbative expansions of (\ref{overlap 2}) up to quadratic terms, we obtain
\ba
&& 1-\braket{ \tilde{\Omega}_2(\epsilon)  | \Omega_1 } \no
&& =\dfrac{1}{2}  \int_\epsilon^{\infty} d\tau \int_{-\infty}^{-\epsilon} d\tau'
\int d^dx \int d^dx' \la  \delta \mathcal{L} (\tau , x)  \delta \mathcal{L} (\tau' , x') \lb, \nonumber
\ea
where we assumed the time reversal symmetry relation $\la\delta \mathcal{L} (-\tau , x)  \delta \mathcal{L} (-\tau' , x')\lb=\la  \delta \mathcal{L} (\tau , x)  \delta \mathcal{L} (\tau' , x') \lb$.

In this way, the information metric with respect to the $\lambda$ perturbation (\ref{pert}) is computed as
\be
G_{\lambda\lambda}=\dfrac{1}{2}\!  \int_\epsilon^{\infty} d\tau\! \int_{-\infty}^{-\epsilon}\! d\tau'
\int d^dx\! \int d^dx'\! \la O(\tau , x)O(\tau' , x') \lb.\label{overlapleading}
\ee
Note that up to now, our argument can be applied to any local operator $O$ in any
quantum field theory. However, for simplicity, we would like to focus on the case where the spin of the primary field $O$ is zero and its total conformal dimension is $\Delta$ in this letter. We can generalize our analysis to a current or an energy stress tensor as we present the details in appendix A and B.

The (normalized) two point function of the primary field takes the universal form
\begin{equation}
\la O(\tau , x) O(\tau', x') \lb = \dfrac{1}{ ((\tau-\tau')^2 +(x-x')^{2} )^{\Delta} }.
\label{correlation d+1}
\end{equation}
By plugging (\ref{correlation d+1}) into
(\ref{overlapleading}), when $d+2-2\Delta<0$ we obtain
\be
G_{\lambda\lambda}=N_d\cdot V_d\cdot \epsilon^{d+2-2\Delta}, \label{cfti}
\ee
where we define $N_d=\f{2^{d-2\Delta}\pi^{d/2}\Gamma(\Delta-d/2-1)}
{(2\Delta-d-1)\Gamma(\Delta)}$ and $V_d$ is the infinite volume of $R^d$. In particular, for a marginal perturbation we have $d+2-2\Delta=-d$. On the other hand, when $d+2-2\Delta\geq 0$, there exists an infrared (IR) divergence and we need a IR cut off $L$ for both the $\tau$ and $x$ integral. This leads to $G_{\lambda\lambda}\propto V_d\cdot L^{d+2-2\Delta}$ (agreeing with \cite{Fid}), where for $d+2-2\Delta=0$
we regard $L^{0}$ as $\log(L/\ep)$.\\

{\bf Holographic Computation}

Now we would like to turn to holographic calculations. For this we focus on the case where the
perturbation (\ref{pert}) is exactly marginal $\Delta=d+1$.
This greatly simplifies the computation in the gravity dual. This is because gravity backgrounds dual to both of $\ket{\Omega_i}\ (i=1,2)$ are the pure AdS$_{d+2}$ with the same radius $R$. Such a gravity dual which interpolates two AdS spaces is called a Janus solution \cite{janus}. The massless bulk scalar field dual to the exactly marginal operator $O$ is denoted by $\phi$.

Let us first study the AdS$_3$ Janus solution introduced in \cite{BGH}. This setup is defined by the action:
\be
S=-\f{1}{16\pi G_N}\int dx^3 \s{g}
\left({\cal R} - g^{a b} \partial_a \phi \partial_b \phi +\f{2}{R^2}\right).
\label{janus-acf}
\ee
The Janus solution is given by the metric
\ba
&& ds^2=R^2\left(dy^2+ f(y) ds^2_{AdS2}\right), \no
&& f (y) = \frac{1}{2} (1+\sqrt{1-2\gamma^2} \cosh(2y) ),
\label{eq: janus-metric}
\ea
and the dilaton
\be
\phi (y) = \gamma\int^{y}_{-\infty}\f{dy}{f (y)}+\phi_1, \label{phj}
\ee
where $\gamma~ (\leq \f{1}{\s{2}})$ is the parameter of Janus deformation. The metric of AdS$_2$ slice is given by $ds^2_{AdS2}=(dz^2+dx^2)/z^2$. $\phi_1=\phi(-\infty)$ is dual to the coupling constant of the exactly marginal deformation
for the ground state $\ket{\Omega_1}$. On the other hand, the value $\phi_2=\phi(\infty)$ for the other ground state $\ket{\Omega_2}$ is obtained by performing the integral in (\ref{phj}) as
$\phi_2-\phi_1=\s{2}\arctan\left[\f{1-\s{1-2\gamma^2}}{\s{2}\gamma}\right]\simeq \gamma$ when $\gamma\ll 1$.

 By matching the asymptotic behavior of the metric (\ref{eq: janus-metric}) at the infinity
 $|y| = y_{\infty}(\to \infty)$
with that of undeformed metric ($\gamma = 0$)
\be
ds^2_{pure}=R^2\left(d\hat{y}^2+ \frac{1}{2} (1+\cosh(2\hat{y})) ds^2_{AdS2}\right),
\ee
we find the following condition
\be
\sqrt{1-2\gamma^2} e^{2y_{\infty}} = e^{2\hat{y}_{\infty}}.
\label{eq:cond_infnity}
\ee

The on-shell action of (\ref{janus-acf}) is evaluated by
\ba
S(\gamma)
&&= \f{R}{4\pi G_N} V_{AdS_2} \cdot
\int^{y_\infty}_{-y_\infty} dy \left\{ \frac{1}{2} (1+\sqrt{1-2\gamma^2} \cosh(2y) ) \right\}  \notag \\
&&= \f{R}{4\pi G_N} V_{AdS_2} \cdot
\left\{ y_{\infty} + \frac{1}{2} \sqrt{1-2\gamma^2} \sinh(2y_{\infty}) \right\},
\label{janus-onacf}
\ea
where $V_{AdS_2}=\int dx\int^\infty_\ep \f{dz}{z^2}=\f{V_1}{\ep}$ is the volume of AdS$_2$ with a unit radius.

By using the condition (\ref{eq:cond_infnity}) at the infinity
\ba
S(\gamma)-S(0)
&&= \f{R}{4\pi G_N} V_{AdS_2} \cdot (y_{\infty} - \hat{y}_{\infty}) \notag \\
&&= \f{RV_1}{16\pi G_N \ep}  \cdot
\log \left(\frac{1}{1 - 2 \gamma^2} \right)  > 0.
\ea
For small $\gamma^2$, we finally find
\be
|\braket{\Omega_2|\Omega_1}|=e^{-S(\gamma)-S(0)}
\simeq 1-\f{RV_1}{8\pi G_N \ep}\gamma^2 .
\ee
Therefore the information metric is estimated as follows
\be
G_{\gamma\gamma}=\f{cV_1}{12\pi\ep},
\ee
where we employed the holographic expression of the central charge $c=\f{3R}{2G_N}$.
Since the normalization of scalar field (\ref{janus-acf}) leads to the two point function
of $O$ which is proportional to the central charge $c$ (i.e. $\delta\lambda\propto \s{c}\delta\phi=\s{c}\gamma$) , we indeed obtain the advertised
formula (\ref{volume}), where $\Sigma$ is given by the AdS$_2$ slice $\rho=0$ in AdS$_3$.

In order to study higher dimensional examples in a universal way, we would like to consider a
simple holographic model, which turns out to give an excellent approximation to various explicit examples. This holographic model is obtained by identifying the exactly marginal deformation at the time slice $\tau=0$ in CFT$_{d+1}$ with a $d+1$ dimensional defect brane $\Sigma$ with a tension $T$, which extends from the time slice on the AdS boundary to the bulk. This is similar to the holographic constructions in \cite{KR,AKTT,AdSBCFT}. In the gravity setup this is simply realized by adding the defect brane action
\be
S_{brane}=T\int_{\Sigma}\s{g}, \label{bqc}
\ee
to the Einstein-Hilbert action. This prescription is consistent with the boundary (or defect) conformal symmetry in a way similar to the AdS/BCFT \cite{AdSBCFT}. Again we can describe the perturbation (\ref{pert}) by the profile of a massless scalar field $\phi$.
When the deformation $\delta\lambda$ is infinitesimally small, we have
\be
T\simeq n_d\cdot \f{(\delta\lambda)^2}{R^{d+1}}, \label{bqt}
\ee
where $n_d$ is an $O(1)$ constant which is fixed from the normalization of two point function
(\ref{correlation d+1}). The Einstein equation shows that $T$ is proportional to $(\delta\lambda)^2$, as the bulk stress tensor is quadratic with respect to
the scalar field. The dependence on $R$ can be explained by the dimensional reason or by comparing with the Janus computation. In this limit, we can treat the brane as a probe, ignoring its
back-reaction. Finally we impose the equation of motion with respect to
the brane embedding. This requires that the action (\ref{bqc}) is extremized and thus
in the Lorentzian signature, $\Sigma$ is the maximal area surface $\Sigma_{max}$ which ends on the time slice $\tau=0$ at the AdS boundary. Together with (\ref{bqt}), we reach our claim (\ref{volume}).

For a CFT$_{d+1}$ on $R^{d+1}$, the holographic formula (\ref{volume}) leads to
\ba
G_{\lambda\lambda}&=&n_dV_d\int^\infty_{\ep}\f{dz}{z^{d+1}}=\f{n_dV_d}{d\ep^{d}}, \label{poin}
\ea
which indeed agrees with (\ref{cfti}).

Similarly we can analyze the global AdS$_{d+2}$, which is described by the metric
\be
ds^2=R^2\left(-(r^2+1)dt^2+\f{dr^2}{r^2+1}+r^2d\Omega^2_{d}\right),
\ee
 to obtain the information metric for a CFT$_{d+1}$ on $R\times S^d$.
\be
G_{\lambda\lambda}=n_dV_d\int^{r_{\infty}}_{0}\f{r^d}{\s{r^2+1}}dr =\f{n_dV_d}{d\ep^d}+\ddd,
\label{global}
\ee
where $r_{\infty}\sim 1/\ep$ is dual to the UV cut off of the CFT. It might be curious
that there appears a logarithmic divergent term $\propto \log r_{\infty}$ when $d$ is even, i.e. odd dimensional CFTs. This logarithmic term is analogous to the boundary central charge in BCFT
\cite{NTU}. Also it is clear from (\ref{global}) that $G_{\lambda\lambda}$ is smaller than the flat space one (\ref{poin}) and this is due to the mass gap in CFTs on a compact space.

Another interesting example is the $d+2$ dimensional AdS Schwarzschild black hole
\ba
&& ds^2=R^2\left(\f{dz^2}{h(z)z^2}-\f{h(z)}{z^2}dt^2+\f{\sum_{i=1}^{d}dx_i^2}{z^2}\right),\no
&& h(z)\equiv 1-(z/z_0)^{d+1}, \label{adssoli}
\ea
which is dual to the finite temperature CFT. The parameter $z_0$ is related the temperature $T$ via
$z_0=\f{d+1}{4\pi T}$. The information metric is computed as
\ba
&& G_{\lambda\lambda}=n_dV_d \int^{z_0}_{\ep}\f{dz}{\s{h(z)}z^{d+1}}  \no
&& \ \ \ =\f{n_dV_d}{d} \left(\f{1}{\ep^d}+\f{b_d}{z_0^d}\right), \no
&& b_d\equiv -1+d\int^1_0 dy\left(1-y^{d+1}+\s{1-y^{d+1}}\right)^{-1}  \no
&& \ \ \  = (d-1) \frac{\sqrt{\pi}}{2}  \frac{\Gamma \left(1 + \frac{1}{d+1} \right)}{\Gamma \left(\frac{1}{2} + \frac{1}{d+1} \right) }\geq 0. \label{bhmetq}
\ea
For example, we have $b_1=0,\ b_2\simeq 0.70,\ b_3\simeq 1.31$.

Finally we would like to study a time-dependent example in order to confirm our proposed formula
(\ref{volume}) can be applied to such a non-trivial setup. For this purpose, we consider thermo-field double (TFD) description of finite temperature state in a two dimensional (2d) CFT:
\be
|\Psi_{TFD}(t)\lb\propto e^{-i(H^{(A)}+H^{(B)})t}\sum_{n}e^{-\f{\beta}{4}(H^{(A)}+H^{(B)})}|n\lb_A|n\lb_B, \label{tfd}
\ee
where $H^{(A)}$ and $H^{(B)}$ are the identical Hamiltonians for the first and second CFT of the TFD; the states $|n\lb_{A,B}$ are the unit norm energy eigenstates in the two CFTs. This TFD setup is dual to the extended geometry of eternal AdS black hole depicted in Fig.\ref{fig:pe}.

We are interested in the inner product
$\la\Psi'_{TFD}(t)|\Psi_{TFD}(t)\lb$ and the information metric $G_{\lambda\lambda}$. Here, the state $\la\Psi'_{TFD}(t)|$ is the TFD state with a Hamiltonian $H'^{(A)}+H'^{(B)}$ which is obtained by an infinitesimal exactly marginal $\lambda$-perturbation (\ref{pert})
with respect to each of $H^{(A)}$ and $H^{(B)}$ at the same time $t$. We argue this deformation is dual to introduce a defect brane $\Sigma$ in the BTZ black hole as in Fig.\ref{fig:pe}.

Let us compute $G_{\lambda\lambda}$ in an Euclidean path-integral formalism of a 2d CFT.
The two point function on $S^1\times R$, where $S^1$ is the thermal circle with periodicity $\beta$,
reads
\be
\la O(x_1,\tau_1)O(x_2,\tau_2)\lb\!=\!\f{\left(\f{\pi}{\beta}\right)^{2\Delta}}
{\left(\!\sinh^2\left(\!\f{\pi(x_1-x_2)}{\beta}\!\right)
\!+\!\sin^2\left(\!\f{\pi(\tau_1-\tau_2)}{\beta}\!\right)\!\right)^{\Delta}}. \nonumber
\ee

Then, $G_{\lambda\lambda}$ at the Euclidean time $\tau$ is expressed as
\ba
&& G_{\lambda\lambda} \no
&& =\f{1}{2}\int^\infty_{-\infty}\!\!\!\!\!\! dx_1dx_2 \int^{\f{3\beta}{4}-\tau-\ep}_{\f{\beta}{4}+\tau+\ep}\!\!\!\!\!\! d\tau_2 \int^{\f{\beta}{4}+\tau-\ep}
_{-\f{\beta}{4}-\tau+\ep}\!\!\!\!\!\! d\tau_1 \la O(x_1,\tau_1)O(x_2,\tau_2)\lb.  \no \label{tpttt}
\ea
Using the formula (for $0\leq t \leq \pi$)
\ba
&&\int^\infty_{-\infty}dx \left(\sinh^2 x+\sin^2 t\right)^{-2} \no
&&=\f{1}{\sin^2 t\cdot \cos^2 t}+(t-\pi/2)\cdot \f{2\sin^2 t-1}{\sin^3 t\cos^3 t},
\ea
we can evaluate the information metric (\ref{tpttt}) when $\Delta=2$, up to finite terms in the
$\ep\to 0$ limit:
\ba
G_{\lambda\lambda}=\f{\pi V_1}{8\ep}
\!-\!\f{\pi V_1}{2\beta}\!+\!\f{2\pi^2 V_1}{\beta^2}\cdot\tau \cot\left(\f{4\pi\tau}{\beta}\right).
\ea
 When $\tau=0$, we simply find $G_{\lambda\lambda}=\f{\pi V_1}{8\ep}$. This is consistent with our holographic result $b_1=0$ in (\ref{bhmetq}).

To study the real time evolution, we have only to set $\tau=it$ and we obtain the result:
\ba
G_{\lambda\lambda}=\f{\pi V_1}{8\ep}-\f{\pi V_1}{2\beta}+\f{2\pi^2 V_1}{\beta^2}\cdot \f{\cosh\f{4\pi t}{\beta}}{\sinh\f{4\pi t}{\beta}}\cdot t. \label{xxy}
\ea
At late time $t \gg \beta$, it grows linearly:
\ba
G_{\lambda\lambda}\simeq \f{\pi V_1}{8\ep}+\f{2\pi^2 V_1}{\beta^2}\cdot t. \label{tfdli}
\ea

\begin{figure}
  \centering
  \includegraphics[width=5cm]{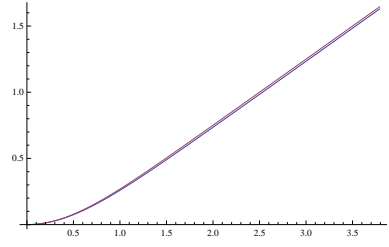}
  \caption{We compare the finite part of our holographic result $\mbox{Vol}(\Sigma)/(R^{d+1}V_1)$ (blue) with that of our CFT result $2G_{\lambda\lambda}/V_1$ (red) as a function of the time $t$ for the TFD state. They deviates only slightly. We set $\beta=2\pi$.}
\label{fig:tfd}
  \end{figure}

We now turn to the holographic computation in the BTZ black hole, where the region (I) in
Fig.\ref{fig:pe} is described by the metric (we set $\beta=2\pi$ for simplicity)
\be
ds^2=R^2(-\sinh^2\rho dt^2+d\rho^2+\cosh^2\rho dx^2).
\ee
We can continue into the region (II) by setting $\kappa=-i\rho$ and $\ti{t}=t+\f{\pi i}{2}$.
In the region (II), if we specify $\Sigma$ by $\kappa=\kappa(\ti{t})$ of $\Sigma$, its volume is given by
\be
\mbox{Vol}(\Sigma)=R^2V_1\int d\ti{t} \cos\kappa\s{\sin^2\kappa-(\de\kappa/\de{\ti{t}})^2}.\label{vol}
\ee
We define $\kappa_*\ \ (0\leq \kappa_*<\pi/4)$ to be the value of $\kappa$ where $\de\kappa/\de{\ti{t}}=0$. We can maximize the volume (\ref{vol}) and extend the solution into the region (I) in a way similar to \cite{HaMa}.
In the end we obtain the following expression of $\mbox{Vol}(\Sigma_{max})$ as a function of $t$
in terms of the parameter $\kappa_*$:
\ba
&&\f{\mbox{Vol}(\Sigma)}{R^{d+1}V_1}=2\sinh\rho_{\infty}+2\int^{\kappa_*}_0
d\kappa\f{\cos\kappa}{\s{\sin^2 (2\kappa_*)/\sin^2 (2\kappa)-1}} \no
&&-2\int^{\rho_{\infty}}_0 d\rho \f{\cosh\rho\left(\s{\sinh^2(2\rho)+\sin^2 (2\kappa_*)}-\sinh (2\rho)\right)}
{\s{\sinh^2(2\rho)+\sin^2 (2\kappa_*)}},\no
&& t=\int^{\kappa_*}_0
\f{d\kappa}{\sin\kappa\s{1-\sin^2 (2\kappa)/\sin^2 (2\kappa^*)}}\no
&&-\int^{\rho_{\infty}}_0 \f{d\rho}{\sinh\rho\s{1+\sinh^2 (2\rho)/\sin^2 (2\kappa^*)}},
\ea
where $\rho_{\infty}$ is the UV cut off such that $e^{\rho_{\infty}}\propto 1/\ep$.
In the late time limit (i.e. $\kappa_*\to \pi/4$) we find the finite part
$\mbox{Vol}_{finite}(\Sigma)/R^{d+1}$ approaches to $V_1\cdot t$, which agrees with
$2G_{\lambda\lambda}$ in (\ref{tfdli}). This linear $t$ growth clearly comes from the Einstein-Rosen bridge as already noted in \cite{Su}. In Fig.\ref{fig:tfd} we plotted the holographic
result versus the CFT result, which shows only a very small deviation.
For example, in the limit $t\to 0$ (or $\kappa_*\to 0$) we find $\mbox{Vol}_{finite}(\Sigma)/(R^{d+1}V_1)\simeq \f{2}{\pi}t^2$, while $2G_{\lambda\lambda}/V_1
\simeq \f{2}{3}t^2$.

In this letter, we introduced an information metric in CFTs and proposed its holographic formula
based on a simple probe model. We presented a number of evidences which support our proposal.
It might be interesting to note that if we normalize the two point
function such that it is proportional to a central charge or if we employ some combination of energy stress tensors (related to some metric perturbations), we have
the estimation $G_{\lambda\lambda}\sim \f{\mbox{Vol}(\Sigma_{max})}{G_N R}$, which is
the same formula argued in \cite{Su} to measure the amount
of complexity.\\

{\bf Acknowledgements} We are very grateful to Adam Brown and Leonard Susskind for careful reading of the draft of this paper and giving us valuable comments. We also thank David Berenstein, Andreas Karch, Shunji Matsuura, Masahiro Nozaki, Shinsei Ryu, Kostas Skenderis, Yu Watanabe, Toby Wiseman and Beni Yoshida for useful comments. We are also thankful to the organizers and participants of the workshop
``Numerical approaches to the holographic principle, quantum gravity and cosmology'', held in YITP, Kyoto University, where the results in this paper have been presented. TT is grateful to the organizers and participants of the KITP program
 ``Entanglement in Strongly-Correlated Quantum Matter,'' held in KITP, UCSB, where important progresses in this paper have been made. TN, NS and KW are supported by JSPS fellowships. TT is supported by JSPS Grant-in-Aid for Scientific Research (B) No.25287058. TT is also supported by World Premier International Research Center Initiative (WPI Initiative) from the Japan Ministry of Education, Culture, Sports, Science and Technology
(MEXT).


\appendix

{\bf Appendix A: Information Metric for Gauge Field Perturbations in CFTs}

We consider a CFT perturbed by the conserved current,
\begin{equation}
S_{tot}=S_{CFT}+\int d^d x d \tau A_{\mu}(x)J^{\mu}(\tau,x),
\end{equation}
where $\partial_{\mu} J^{\mu}=0$.
In this case the overlap in the leading order of the perturbation is given by
\begin{equation}
1-\braket{ \tilde{\Omega}_2(\epsilon)  | \Omega_1 }
=\dfrac{1}{2} \int d^dx \int d^dy \delta A^{\mu}(x)\delta A^{\nu}(y) J_{\mu \nu}(x,y)
\end{equation}
where we have defined
\begin{equation}
\begin{split}
&J_{\mu \nu}(x-y) \equiv \int_\epsilon^{\infty} d\tau \int_{-\infty}^{-\epsilon} d\tau'
\langle J_{\mu}(\tau,x) J_{\nu}(\tau' ,y) \rangle  \\
& =  \int_{2 \epsilon}^{\infty} d (\tau-\tau')((\tau-\tau') -2 \epsilon)
\langle J_{\mu}(\tau,x) J_{\nu}(\tau' ,y) \rangle     .
\end{split}
\end{equation}
We can calculate $J_{\mu \nu}$ by using the explicit expression for
the correlation function of the conserved currents \cite{Obsorn:1993},
\begin{equation}
\begin{split}
\langle J_{\mu}(\tau,x) J_{\nu}(\tau' ,y) \rangle
&=\dfrac{C_V}{((\tau-\tau')^2+(x-y)^2)^{d}} \\
& \times \left( \delta_{\mu \nu}-2\dfrac{(x-y)_{\mu} (x-y)_{\nu} }{(\tau-\tau')^2+(x-y)^2} \right) .
\end{split}
\end{equation}
where $C_V$ is a constant determining the overall scale of this two point function.

For $d>1$, we obtain
\begin{equation}
\begin{split}
&J_{\tau \tau}(x)/C_V \\
&=  \left(4 \epsilon^2+r^2\right)^{-d} \left[-4 \epsilon^2 \left(\frac{4
   \epsilon^2}{r^2}+1\right)^d \, _2F_1\left(\frac{1}{2},d;\frac{3}{2};-\frac{4
   \epsilon^2}{r^2}\right)  \right. \\
 & \left.  +8 \epsilon^2 \,
   _2F_1\left(1,\frac{1}{2}-d;\frac{3}{2};-\frac{4
   \epsilon^2}{r^2}\right)+\frac{(d-2) r^2-4 \epsilon^2 d}{2 (d-1)
   d}\right]  \\
 &-  \frac{\sqrt{\pi } \epsilon (d-1)
   \left(\frac{1}{r^2}\right)^{d-\frac{1}{2}} \Gamma
   \left(d-\frac{1}{2}\right)}{\Gamma (d+1)} ,
\end{split}
\end{equation}
\begin{equation}
\begin{split}
&J_{\tau j}(x)/C_V = -2 x_j \cdot \frac{2^{-2 d-1} \left(4 \epsilon^2+r^2\right)^{-d} }{(2 d-1) r^2 \Gamma (d+1)^2}  \\
& \times \left[\pi  r^4
   \left(\frac{1}{r^2}\right)^{d+\frac{1}{2}} \Gamma (2 d+1) \left(4
   \epsilon^2+r^2\right)^d   \right. \\
& \left.  +\frac{4^{d-2} r^2 \left(4 \epsilon^2+r^2\right) \Gamma (d-1)
   \Gamma (d+1) }{\epsilon^3} \right. \\
& \left. \times  \left(\left(4 \epsilon^2+r^2\right) \,
   _2F_1\left(1,\frac{1}{2}-d;-\frac{1}{2};-\frac{4 \epsilon^2}{r^2}\right)  \right. \right. \\
& \left. \left.    +8
   \epsilon^2 (d-1)-r^2\right) \right]  ,
\end{split}
\end{equation}
and
\begin{equation}
\begin{split}
&J_{ i j }(x)/C_V= \delta_{i j } \\
&\times  \left[
\frac{ \left(8 \epsilon^2 (d-1) \,
   _2F_1\left(1,\frac{3}{2}-d;\frac{3}{2};-\frac{4
   \epsilon^2}{r^2}\right)+r^2\right)}{2 (d-1) r^2 \left(4 \epsilon^2+r^2\right)^{d-1}} \right.  \\
 & \left.  -\frac{\sqrt{\pi } \epsilon
   \left(\frac{1}{r^2}\right)^{d-\frac{1}{2}} \Gamma
   \left(d-\frac{1}{2}\right)}{\Gamma (d)}  \right]   -2 x_i x_j  \\
& \times \left[
\frac{ \left(8 \epsilon^2 d \,
   _2F_1\left(1,\frac{1}{2}-d;\frac{3}{2};-\frac{4
   \epsilon^2}{r^2}\right)+r^2\right)}{2 d r^2 \left(4 \epsilon^2+r^2\right)^{d}}   \right. \\
& \left.   -\frac{\sqrt{\pi } \epsilon
   \left(\frac{1}{r^2}\right)^{d+\frac{1}{2}} \Gamma
   \left(d+\frac{1}{2}\right)}{\Gamma (d+1)}   \right] ,
\end{split}
\end{equation}
where $r\equiv ||x||$.

For $d=1$, we need an IR cutoff for the integrals in $J_{\mu \nu}$.
In order to reguralize the integrals,
we replace
$  \int_\epsilon^{\infty} d\tau \int_{-\infty}^{-\epsilon} d\tau'
= \int_{2 \epsilon}^{\infty} d (\tau-\tau')((\tau-\tau') -2 \epsilon)$
with $\int_{2 \epsilon}^{T} d (\tau-\tau')((\tau-\tau') -2 \epsilon)$ ,
where $T$ is an  IR cutoff length.
Thus we obtain
\begin{equation}
\begin{split}
&J_{\tau \tau}(x)/C_V= -J_{x x}(x)/C_V \\
& = \dfrac{T(-2\epsilon+T)}{x^2+T^2} +\dfrac{1}{2} \ln \left( \dfrac{4 \epsilon^2+x^2}{x^2+T^2} \right) ,
\end{split}
\end{equation}
and
\begin{equation}
\begin{split}
&J_{\tau x}(x)/C_V =
 \arctan \left( \dfrac{2\epsilon}{x} \right) -\arctan \left( \dfrac{T}{x} \right) -\dfrac{x(2\epsilon-T)}{x^2+T^2} .
\end{split}
\end{equation}
It is useful to perform the Fourier transformation:
$\tilde{J}_{\mu \nu}(k) \equiv \int_{-\infty}^{\infty} dx e^{-ikx} J_{\mu \nu} (x) $,
which leads to
\begin{equation}
\begin{split}
&\tilde{J}_{\tau \tau}(k)/C_V= -\tilde{J}_{x x}(k)/C_V \\
&= \pi \left[ \dfrac{1}{|k|} \left( e^{-T |k|} -e^{-2 \epsilon |k|} \right) + \right. \\
& \left.
 (T-2\epsilon) \left( e^{k T} \theta (-k) + e^{-k T} \theta (k) \right) \right]  \\
& \xrightarrow{T\to \infty, \epsilon \to 0} - \dfrac{\pi}{|k|}
\end{split}
\end{equation}
and
\begin{equation}
\begin{split}
&\tilde{J}_{\tau x}(k)/C_V= \\
&-i \pi \left[ \dfrac{1}{k} \left( e^{-T |k|} -e^{-2 \epsilon |k|} \right) + \right. \\
& \left.  (2\epsilon-T) \left( e^{k T} \theta (-k) - e^{-k T} \theta (k) \right) \right] \\
& \xrightarrow{T\to \infty, \epsilon \to 0}  \dfrac{i \pi}{k}  .
\end{split}
\end{equation}
After we go back to the original Lorentz signature by the Wick rotation $\tau=it$,
 for $T\to \infty$ and  $ \epsilon \to 0$, we obtain
\begin{equation}
\begin{split}
&\tilde{J}_{t t}(k)/C_V= \tilde{J}_{x x}(k)/C_V= \dfrac{\pi}{|k|} , ~~~
 \tilde{J}_{t x}(k)/C_V= \dfrac{\pi}{k} .
\end{split}
\end{equation}
Thus the eigenvalues of $ \tilde{J}_{\mu \nu} /C_V $ are $ \left( \frac{\pi}{|k|} \pm \frac{\pi}{k} \right) \geq 0 $,
and the information metric is non-negative.\\

{\bf Appendix B: Information Metric for Metric Perturbations in CFTs}

In this section, we consider a CFT perturbed by the energy momentum tensor:
\be
S_{tot}=S_{CFT}+\int d^{d}xdt~g_{\mu\nu}(x)T^{\mu\nu}(\tau , x)
\ee
In this case the overlap in leading order of the perturbation is given by
\be
\braket{ \tilde{\Omega}_2(\epsilon)  | \Omega_1 }=1-\f{1}{2}\int d^{d}x d^{d}y~\delta g^{\mu\nu}(x)
\delta g^{\sigma\rho}(y)Q_{\mu\nu\sigma\rho}(x-y)
\ee
where we defined
\be
Q_{\mu\nu\sigma\rho}(x-y)=\int^\infty_\ep dt\int^{-\ep}_{-\infty} dt'\la T_{\mu\nu}(x,t)T_{\sigma\rho}(y,t')\lb.
\ee
We can calculate $Q_{\mu \nu\sigma\rho}$ by using the explicit expression for
the correlation functions of the energy momentum tensor \cite{Obsorn:1993},
\ba
&& \la T_{\mu\nu}(x,t)T_{\sigma\rho}(y,t')\lb \no
&& =\f{C_{T}\mathcal{E}^{T}_{\alpha\beta:\sigma\rho}}
{((x-y)^2+(t-t')^2)^{d+1}}R_{\mu\alpha}(x,t:y,t')R_{\nu\beta}(x,t;y,t').\no
\ea
where
\be
\mathcal{E}^{T}_{\mu\nu:\eta\xi}=\f{1}{2}(\delta_{\mu\eta}\delta_{\nu\xi}+\delta_{\mu\xi}\delta_{\nu\eta})-\f{1}{d+1}\delta_{\mu\nu}\delta_{\xi\eta},
\ee
\be
R_{\mu\nu}(x,t:y,t')=\delta_{\mu\nu}-2\f{(x-y)_{\mu}(x-y)_{\nu}}{(x-y)^2+(t-t')^2},
\ee
and $C_{T}$ is a constant determining the overall scale of this two point function, which is related to the central charge of the CFT.

In real space coordinates, we obtain
\ba
&&Q_{tttt}(x)/C_{T}=\frac{\left(r^2+4 \epsilon ^2\right)^{-d-1}}{2(d+1)(d+2) }\no
&&
\times\Bigg[ (-2+d)r^2+4(-2+d+2d^2)\epsilon^2  
\no&&
\left.-8(-1+d)(1+d)^{2}\epsilon ^2\f{{}_{2}F_{1}\left(-\f{1}{2}, 1; \f{3}{2}+d; -\f{r^2}{4\epsilon ^2}\right)}{d+\f{1}{2}}\right] 
\no&& Q_{x^{i}ttt}(x)/C_{T}
= x^{i}
\frac{\left(r^2+4 \epsilon ^2\right)^{-d-2}}{16 \epsilon^3\Gamma (d+3)}
\no&&\times\Bigg[32\epsilon^{4}(dr^2-4(2+d)\epsilon^2)\Gamma (d+1) \no&&
+16^{-d}(1+d)(2+d)\sqrt{\pi}\epsilon^{-2d}\Gamma(1+2d)(r^2+4\epsilon^2)^{2+d}\no&&
\times\bigg(8\epsilon^{2}\f{{}_{2}F_{1}\left(d+\f{1}{2}, 3+d; \f{3}{2}+d; -\f{r^2}{4\epsilon ^2}\right)}{\f{3}{2}+d}
\no&&\left.-(1+2d)r^2\f{{}_{2}F_{1}\left(d+\f{3}{2}, 3+d; \f{5}{2}+d; -\f{r^2}{4\epsilon ^2}\right)}{\f{5}{2}+d}\bigg)\right],
\no&&Q_{x^{i}tx^{j}t}( x)/C_{T}
=\delta_{ij}
\frac{-\left(r^2+4 \epsilon ^2\right)^{-d-1}}{4\Gamma (d+2)}
\no&&\times\Bigg[(r^2+4\epsilon^2)\Gamma (d)-(r^2-4\epsilon^2)\Gamma (d+1) \no&&
+4^{-d}\epsilon^{-2d}(r^2+4\epsilon^2)^{1+d}\Gamma(2+d)\no&&
\times\bigg(-2\f{{}_{2}F_{1}\left(d+\f{1}{2}, 2+d; \f{3}{2}+d; -\f{r^2}{4\epsilon ^2}\right)}{1+2d}
\no&&\left.+r^2\f{{}_{2}F_{1}\left(d+\f{3}{2}, 2+d; \f{5}{2}+d; -\f{r^2}{4\epsilon ^2}\right)}{2(3+2d)\epsilon^2}\bigg)\right]
\no&&
   +x^{i}x^{j}
   \frac{1}{16\Gamma (d+3)}\Bigg[24(r^2+4(2+d)\epsilon^2)\Gamma(1+d)\no&&
   -8r^2(r^2+4\epsilon^2)^{-d-2}\Gamma (d+2) +4^{-d}\epsilon^{-2d-4}\Gamma(3+d)\no&&
\times\bigg(-12\epsilon^2\f{{}_{2}F_{1}\left(d+\f{3}{2}, 3+d; \f{5}{2}+d; -\f{r^2}{4\epsilon ^2}\right)}{3+2d}
\no&&\left.+r^2\f{{}_{2}F_{1}\left(d+\f{5}{2}, 3+d; \f{7}{2}+d; -\f{r^2}{4\epsilon ^2}\right)}{5+2d}\bigg)\right],
\ea
\ba
&&Q_{x^{i}x^{j}tt}( x)/C_T\no
&&=-\f{1}{d+1}\delta_{ij}\Bigg(\frac{\left(r^2+4 \epsilon ^2\right)^{1-d}}{16
   (d-1) d r^2 \epsilon ^2} \no
 &&\left(r^2-\left(r^2+4 \epsilon ^2\right) \,
   _2F_1\left(1,\frac{1}{2}-d;-\frac{1}{2};-\frac{4 \epsilon ^2}{r^2}\right)\right)\no
 &&-\frac{\sqrt{\pi } \epsilon  r^{-2 d-1} \Gamma
   \left(d+\frac{1}{2}\right)}{d!}\Bigg) \no
&&\ \ \ + 4 x^{i}x^{j}\Bigg(\frac{\left(r^2+4 \epsilon ^2\right)^{-d-1}}{16 (d+1) (d+2) (2 d+3) \epsilon ^2} \no
&&\Bigg(-\left(r^2+4 \epsilon ^2\right) \,
   _2F_1\left(1,-d-\frac{3}{2};-\frac{1}{2};-\frac{4 \epsilon ^2}{r^2}\right)\no
 &&+8 (d+2)
   \epsilon ^2+r^2\Big)-\frac{\sqrt{\pi }
   \epsilon  r^{-2 d-3} \Gamma \left(d+\frac{3}{2}\right)}{2 \Gamma (d+3)} \Bigg)
\no
\ea
\ba
&&Q_{x^{i}x^{j}x^{k}t}( x)/C_T\no
&&=-( x^{i}\delta_{jk}+ x^{j}\delta_{ik})\f{1}{32 d (d+1) (2
   d+1) \epsilon ^3 \Gamma (d+2)}\no
&&\Bigg(r^{-2 d-1} \left(r^2+4 \epsilon ^2\right)^{-d} \Bigg(r^{2 d+1}
\Gamma (d+2) \no
&&\left(\left(r^2+4 \epsilon ^2\right) \,
   _2F_1\left(1,-d-\frac{1}{2};-\frac{1}{2};-\frac{4 \epsilon
^2}{r^2}\right)+8 d
   \epsilon ^2-r^2\right)\no
 &&+16 \sqrt{\pi } d (d+1) \epsilon ^3 \Gamma
   \left(d+\frac{3}{2}\right) \left(r^2+4 \epsilon
^2\right)^d\Bigg)\Bigg) \no
&&+4 x^{i}x^{j}x^{k}\frac{1}{32 (d+1)
   (d+2) (2 d+3) \epsilon ^3 \Gamma (d+3)}\no
&&\Bigg(r^{-2 d-3} \left(r^2+4 \epsilon ^2\right)^{-d-1} \Bigg(r^{2 d+3}
\Gamma (d+3) \no
&&\left(\left(r^2+4 \epsilon ^2\right) \,
   _2F_1\left(1,-d-\frac{3}{2};-\frac{1}{2};-\frac{4 \epsilon
^2}{r^2}\right)+8 (d+1)
   \epsilon ^2-r^2\right)\no
&&+16 \sqrt{\pi } (d+1) (d+2) \epsilon ^3 \Gamma
   \left(d+\frac{5}{2}\right) \left(r^2+4 \epsilon
^2\right)^{d+1}\Bigg) \Bigg)\no
\ea
\ba
&&
Q_{x^{i}x^{j}x^{k}x^{l}}({\bf x})/C_T\no
&&=(\f{1}{2}\Big(\delta_{ik}\delta_{jl}+\delta_{il}\delta_{jk})-\f{1}{d+1}\delta_{ij}\delta_{kl}\Big)\frac{\left(r^2+4 \epsilon ^2\right)^{1-d}}{16(d-1) d r^2 \epsilon ^2} \no
&&\left(r^2-\left(r^2+4 \epsilon ^2\right) \,
   _2F_1\left(1,\frac{1}{2}-d;-\frac{1}{2};-\frac{4 \epsilon
^2}{r^2}\right)\right)\no
&&-\frac{\sqrt{\pi } \epsilon  r^{-2 d-1} \Gamma
   \left(d+\frac{1}{2}\right)}{d!}
 \no
&&-( x^i x^k \delta_{jl}+ x^i x^l \delta_{jk}+ x^j x^l \delta_{ik}+ x^i x^l \delta_{jk})\frac{\left(r^2+4 \epsilon ^2\right)^{-d}}{16 d (d+1) r^2 \epsilon ^2}\no
&&\left(r^2-\left(r^2+4 \epsilon ^2\right) \,
   _2F_1\left(1,-d-\frac{1}{2};-\frac{1}{2};-\frac{4 \epsilon
   ^2}{r^2}\right)\right)\no
 &&-\frac{\sqrt{\pi
} \epsilon
   x^{-2 (d+1)-1} \Gamma \left(d+\frac{3}{2}\right)}{(d+1)!} \no
&&+4 x^ix^jx^kx^l\frac{\left(r^2+4 \epsilon ^2\right)^{-d-1}}{16 (d+1) (d+2) x^2 \epsilon
^2} \no
&&\left(r^2-\left(r^2+4
\epsilon ^2\right) \,
   _2F_1\left(1,-d-\frac{3}{2};-\frac{1}{2};-\frac{4 \epsilon
   ^2}{r^2}\right)\right)\no
&&-\frac{\sqrt{\pi } \epsilon
   r^{-2 (d+2)-1} \Gamma \left(d+\frac{5}{2}\right)}{(d+2)!}\no
\ea

When only $g_{00}$ component is perturbed by a constant, the perturbed Hamiltonian is a constant multiple of the original Hamiltonian. Therefore the information metric for such deformation should be zero identically. We can explicitly confirm this fact by a integration $\int d^{d}{\bf x}~Q_{0000}({\bf x})=0$.

\end{document}